# Beyond Screen Time: Explaining Cross-National Differences in Digital Literacy through Socioeconomic and Psychological Mechanisms

Hyejeong Lee
The University of Texas at San Antonio
hyejeong.lee@utsa.edu

Daeyoung Ham
The University of Texas at San Antonio
daeyoung.ham@utsa.edu

Suyoun Kim
Pusan National University
suyoun2225@pusan.ac.kr

Tiffany Emanuel
The University of Texas at San Antonio
tiffany.emanuel@my.utsa.edu

**Abstract**

This study provides a structural explanation for cross-national variation in the relationship between screen time and digital outcomes. While prior research and large-scale assessments such as ICILS have documented inconsistent associations between screen time and digital competence, the mechanisms underlying these differences remain unclear. Using ICILS 2023 data, this study employs multigroup structural equation modeling to examine the relationships among socioeconomic status, screen time regulation, ICT self-efficacy, and digital literacy outcomes. Results reveal substantial cross-country differences in the effects of screen time regulation. In contrast, ICT self-efficacy emerges as a consistent and robust predictor across all countries. Moreover, screen time regulation influences outcomes indirectly through self-efficacy in some contexts but not others. These findings challenge the use of screen time as a standalone indicator of digital engagement and highlight the importance of psychological mechanisms. By integrating socioeconomic, behavioral, and psychological factors, this study advances a more nuanced understanding of digital competence and moves beyond quantity-based approaches to digital learning.

## 1. Background

In contemporary digital societies, students' ability to effectively navigate, evaluate, and engage with digital technologies has become a foundational competence for both academic success and lifelong learning. As educational systems increasingly integrate digital tools into teaching and learning, digital literacy has emerged as a critical indicator of students' readiness to participate in knowledge-based economies (Fraillon et al., 2020; Siddiq et al., 2017). Despite the growing importance of digital competence, substantial disparities persist in students' digital learning opportunities and outcomes. These disparities are often shaped by broader structural factors, including socioeconomic status (SES), access to technological resources, and variations in educational support (van Deursen & van Dijk, 2019; Warschauer & Matuchniak, 2010). A substantial body of research has consistently identified SES as a key determinant of digital literacy, with students from lower-SES backgrounds tending to demonstrate lower levels of digital competence (Scherer & Siddiq, 2019). These inequalities are not limited to access alone, but extend to differences in how students engage with technology, the quality of digital experiences available to them, and the extent to which they are supported in developing digital skills.

Within this broader context, students' patterns of digital engagement, particularly screen time, have received increasing attention as a potential explanatory factor. Screen time is frequently used as a proxy for students' exposure to digital technologies, and is often examined in relation to learning outcomes and well-being. In educational and policy discussions, limiting device use during school days has been framed as an important mechanism for promoting healthy technology habits (Hatlevik et al., 2018). However, the relationship between screen time and digital literacy remains highly ambiguous. While limiting screen time may reduce excessive or non-productive use, it may also restrict opportunities for independent exploration, practice, and skill development in digital environments.

A recent executive summary from the International Computer and Information Literacy Study (ICILS) 2023 highlights this ambiguity (Fraillon, 2025). The international report documents substantial cross-national variation in the relationship between screen time regulation and digital outcomes and notes that the reasons for these cross-national differences remain unclear. This gap is further echoed in prior literature, which suggests that relying on screen time as a primary indicator of digital engagement may be insufficient for explaining students' learning outcomes (Livingstone & Helsper, 2007; Rodriguez-Ayllon et al., 2020) and digital device use alone often fails to predict academic success (Lee & Kim, 2025; Juuti et al., 2022).

These considerations point to an important limitation in existing research. Much of the literature has examined screen time, SES, and self-efficacy as separate predictors of digital outcomes, often relying on bivariate or isolated analyses. However, such approaches may overlook the complex, interdependent relationships among these factors. In particular, the lack of a structural perspective limits our ability to explain why the effects of screen time differ across contexts. Without accounting for how screen time is embedded within broader systems of influence, including socioeconomic conditions and psychological factors, it remains difficult to interpret its role in shaping digital competence.

Accordingly, the present study seeks to address this gap by examining the structural relationships among SES, screen time, ICT self-efficacy, and digital outcomes within a multigroup framework. Thus, this study aims to move beyond descriptive comparisons and provide a theoretically grounded explanation for cross-national variation in the effects of screen time. Specifically, the study investigates whether the inconsistent associations between screen time and digital literacy can be better understood when these variables are modeled simultaneously within a structural equation framework. To guide this investigation, the study addresses the following research questions:

RQ1. How do SES and household screen time regulation directly predict self-efficacy and digital competencies?

RQ2. To what extent does self-efficacy mediate the relationships between SES, screen time regulation, and digital competencies?

RQ3. To what extent do the direct and indirect structural pathways differ across countries?

By addressing these questions, this study contributes to a more nuanced understanding of digital literacy by shifting the focus from quantity-based measures of digital exposure toward the structural and psychological mechanisms that shape students' engagement with technology.

## 2. Literature Review

### 2.1. Socioeconomic Foundations and Digital Literacy

Digital literacy encompasses a broad range of competencies related to accessing, evaluating, creating, and communicating information using digital technologies (Fraillon et al., 2020; UNESCO, 2018). Prior research has identified SES as a structural determinant of digital literacy outcomes (Lee et al., 2025). A meta-analysis revealed a weak but consistent association between SES and digital literacy, with students from higher SES backgrounds outperforming their lower SES peers (Scherer & Siddiq, 2019). Further, recent research further suggests that home digital resources and parental mediation function as parallel mediators linking family SES to digital literacy development (Cao et al., 2025). Families with higher SES provide greater access to digital tools, enriched learning environments, and more structured guidance, all of which contribute to stronger digital competencies.

From an ecological perspective (Bronfenbrenner, 1979), children's digital engagement is shaped by multiple contextual layers, including family resources, parental educational attainment, and neighborhood characteristics. Studies show that children from lower-SES households are more likely to experience higher overall screen exposure but have fewer opportunities for structured or educationally meaningful digital engagement (Carson et al., 2010; MacLeod et al., 2008; Lee et al., 2022). These disparities indicate that socioeconomic inequalities extend beyond access to devices to differences in the quality and purpose of digital use. Students from higher-SES backgrounds are more likely to encounter high-quality digital learning experiences from an early age and to benefit from structured, purposeful, and sustained technology integration across home and school settings, potentially reinforcing existing educational advantages (Cao et al., 2025; Lee et al., 2023). Thus, SES forms a foundational layer through which digital literacy opportunities and outcomes are structured.

Despite consistent evidence linking SES to digital literacy, relatively few studies have examined how these structural inequalities operate differently across national contexts using performance-based measures in technology. Understanding whether and how socioeconomic disparities manifest across diverse educational systems is essential for clarifying the contextual conditions under which digital inequalities are reproduced or mitigated.

### 2.2. Screen Time Limits and Digital Outcomes

Students' digital engagement is frequently operationalized through proxy measures such as screen time and parental restriction practices, which capture distinct dimensions of youth-technology interactions. In both research and policy contexts, screen time has been widely utilized as an indicator of digital exposure, typically examined in relation to academic performance, well-being, and digital skills (Adelantado-Renau et al., 2019). However, contemporary literature highlights that empirical findings regarding the relationship between screen time and learning outcomes remain highly fragmented and inconsistent (Orben & Przybylski, 2019). Some studies suggest that moderate levels of screen use are not associated with adverse developmental or academic outcomes, indicating that screen engagement within normative ranges may be developmentally benign (Ferguson, 2017). In contrast, other research reports that excessive screen time may exert negative influences on students' academic achievement and digital competency, often due to the displacement of learning activities and reduced attention spans (Adelantado-Renau et al., 2019). Consequently, these mixed findings suggest that evaluating the sheer volume of screen time alone is insufficient for capturing its complex educational implications.

To address these quantitative limitations, researchers have increasingly conceptualized students' digital exposure within broader parental mediation frameworks, which differentiate between restrictive and active forms of regulation (Chen & Chang, 2016; Zhao et al., 2023). These mediation strategies yield fundamentally different implications for cognitive and skill development. Restrictive mediation, which primarily involves enforcing rigid limits on device duration or content, has proven effective in mitigating immediate online risks and problematic internet use (Liu et al., 2023; Ren et al., 2022). However, it simultaneously risks, constraining opportunities for the autonomous exploration, trial-and-error, and digital problem-solving necessary to build advanced technological competence (Rodríguez-de-Dios, 2018). In contrast, active mediation, characterized by open discussion, co-use, and guided interaction, fosters purposeful and reflective digital engagement. By facilitating autonomous mastery experiences, active digital parenting is more consistently associated with enhanced digital competence, self-regulation, and self-efficacy (Beyens et al., 2022; Tan et al., 2025; Zhao et al., 2023).

Crucially, while these digital practices manifest at the individual households, they are often embedded within distinct regulatory environments and societal norms. National frameworks governing youth digital infrastructure vary dramatically on a global scale. For example, no statutory time limits exist; instead, public health reports reference WHO-aligned recommendations limiting recreational screen time in Portugal (Pizarro et al., 2023). In Taiwan, national frameworks range from statutory mandates with legal penalties for noncompliance (Lee, 2015; Taiwan Ministry of Health and Welfare, 2021). South Korea has shifted from mandatory restrictions, such as the abolished “shutdown law,” toward voluntary prevention programs amid rising adolescent digital engagement (Kim et al., 2025).

These variations are reflected in large-scale international benchmarks. Recent data from ICILS 2023 reveal substantial cross-national heterogeneity in how screen time restrictions relate to students’ digital outcomes (Fraillon, 2025). In certain educational systems, students navigating unrestrictive home environments demonstrate superior digital literacy, whereas in other countries, the inverse or statistically non-significant patterns emerge. Critically, these international findings remain largely descriptive, leaving the underlying psychological and contextual mechanisms explaining why the effects of screen limits diverge across nations unexplained. Therefore, a rigorous, multi-group examination is warranted to untangle how micro-household regulations and psychological indicators, such as self-efficacy, interact to shape digital competencies across distinct national contexts.

### 2.3. Self-Efficacy as a Psychological Pathway to Digital Literacy

One promising explanation for these inconsistent findings lies in the role of psychological factors, particularly ICT self-efficacy. Rooted in social cognitive theory, self-efficacy refers to individuals’ beliefs in their ability to successfully perform specific tasks (Bandura, 1997). In digital contexts, ICT self-efficacy reflects students’ confidence in their ability to use digital tools for information processing, communication, and problem-solving.

Prior research has consistently shown that higher levels of ICT self-efficacy are associated with more effective digital engagement and better learning outcomes (Lee & Kim, 2025; Scherer & Siddiq, 2019). Students who feel confident in using digital technologies are more likely to engage in complex tasks, persist in problem-solving, and use digital tools in purposeful ways. Empirical evidence further suggests that digital literacy proficiencies and prior technological experiences are closely linked to the development of self-efficacy beliefs (Kim & Lee, 2025; Prior et al., 2016; Yang & Cheng, 2009), and that ICT self-efficacy is a significant predictor of students’ digital performance (Hatlevik et al., 2018).

ICT self-efficacy is also shaped by students’ socioeconomic backgrounds. Existing studies indicate that parental education and occupational status positively predict children’s self-efficacy beliefs (Han et al., 2015; Yang & Chen, 2009). In this sense, self-efficacy can be understood as a psychological pathway through which socioeconomic resources translate into differences in digital competence.

Importantly, the development of ICT self-efficacy is closely related to the quality of digital experiences. Prior research suggests that students who engage in meaningful and purposeful uses of technology, particularly within supportive instructional environments, tend to report higher levels of self-efficacy and demonstrate stronger digital competencies (Hatlevik et al., 2018; Prior et al., 2016; Hristov et al., 2025). These findings reinforce the argument that digital competence is shaped not only by access or exposure, but also by the nature of engagement and the opportunities students have to develop confidence through successful experiences.

Self-efficacy is inherently context-dependent (Bandura, 1997; Markus & Kitayama, 1991). Differences in expectations of learner autonomy and contexts may influence how students develop their efficacy beliefs. Additionally, ICT self-efficacy may function not only as a direct predictor of digital outcomes but also as a key mechanism through which contextual and behavioral factors influence learning. Therefore, rather than exerting direct effects, variables such as screen time or parental regulation may shape students' experiences and confidence, which in turn influence their performance. This perspective suggests that understanding digital competence requires attention to the interplay between behavioral, contextual, and psychological factors.

**2.4. Research Gap**

Existing research highlights the importance of SES, screen time, and self-efficacy in shaping digital competence. However, these factors have often been examined in isolation, with limited attention to how they interact within a broader structural framework. As a result, current research provides limited insight into the mechanisms underlying the inconsistent relationship between screen time and digital outcomes.

In particular, the lack of a structural perspective constrains our ability to explain cross-national variation. While ICILS findings clearly demonstrate that the effects of screen time limit differ across countries, existing studies have largely focused on identifying these differences rather than explaining them. Without accounting for how screen time is embedded within systems of socioeconomic and psychological influences, it remains unclear why similar screen time regulations lead to different outcomes across contexts.

To address this gap, there is a need that integrates behavioral, socioeconomic, and psychological factors within a unified analytical framework. By examining the structural relationships among SES, screen time limit, self-efficacy, and digital outcomes, it becomes possible to move beyond descriptive patterns and toward a more nuanced understanding of how digital competence develops across contexts. Accordingly, the present study employs a multigroup structural equation modeling approach to investigate whether cross-national differences in the effects of screen time can be explained by variations in underlying structural relationships. In doing so, it responds directly to the research gap identified in ICILS and contributes to advancing the field beyond simplistic, quantity-based measures of digital engagement.

## 3. Methods

### 3.1. Data Source and Sample

This study draws on data from the International Computer and Information Literacy Study (ICILS) 2023, a large-scale international assessment designed to evaluate eighth-grade students' competencies in digital contexts (Fraillon, 2025). ICILS assesses students' abilities to use digital technologies for information processing, communication, and problem-solving, with a particular focus on two domains: Computer and Information Literacy (CIL) and Computational Thinking (CT). The assessment is administered through a computer-based testing platform and is accompanied by comprehensive background questionnaires that capture students' demographic, behavioral, and attitudinal characteristics.

For the purposes of this study, three countries were selected to enable cross-national comparison: Korea (KR), Taiwan (TW), and Portugal (PT). These countries were chosen due to their contrasting patterns in the relationship between screen time and digital outcomes reported in ICILS 2023, making them particularly suitable for investigating cross-national variation.

The analytic sample consisted of $n$ = 3,355 students in Korea, $n$ = 4,759 students in Taiwan, and $n$ = 3,138 students in Portugal. All analyses were conducted using student-level data provided in the ICILS international database.

### 3.2. Measures

This study examined the structural relationships among four key constructs derived from the ICILS 2023 dataset: SES, screen time regulation, general ICT self-efficacy, and digital outcomes. SES was modeled as a latent construct representing students' family background. Consistent with ICILS documentation, SES was indicated by multiple variables, including parental education, parental occupation, and home literacy resources. These indicators capture both economic and cultural dimensions of students' home environments and are widely used in international large-scale assessments (Fraillon, 2025).

General ICT self-efficacy refers to students' perceived ability to perform a range of digital tasks. In ICILS, this construct is measured through student responses to items assessing confidence in activities such as searching for information online, evaluating the credibility of digital content, and using digital tools for school-related tasks. These items collectively reflect students' confidence in engaging with everyday digital practices.

Screen time regulation was included as an observed categorical variable based on students' self-reports of parental rules regarding screen use on school days. The variable distinguishes among three categories: (1) restrictions that include time spent on schoolwork, (2) restrictions that exclude school-related screen use, and (3) no restrictions. In the analysis, this variable was treated as an ordinal predictor.

Digital outcomes were represented by students' performance in CIL and CT, as measured through the ICILS standardized computer-based assessment. CIL reflects students' ability to investigate, create, and communicate information while CT focuses on students' ability to conceptualize problems and develop algorithmic solutions, reflecting more advanced forms of digital problem-solving. The analyses used the scaled scores provided in the ICILS dataset, which are reported as plausible values to account for measurement uncertainty in large-scale assessments.

**3.3. Analytical Strategy**

This study employed a multigroup structural equation modeling (SEM) approach to examine the relationships among SES, screen time, ICT self-efficacy, and digital outcomes across three countries (KR, PT, TW). The multigroup framework was adopted to allow all structural parameters to be estimated freely across countries, thereby enabling a direct examination of cross-national differences in the relationships among the variables.

The specified model included SES as an exogenous predictor, screen time limits (ST) as an ordered endogenous variable, general ICT self-efficacy (G) as a continuous mediator, and two continuous outcome variables. The structural model simultaneously estimated (a) direct effects of SES and screen time on both outcomes, (b) the effect of SES and screen time on ICT self-efficacy, and (c) the indirect effects of SES and screen time on digital outcomes through ICT self-efficacy. Residual covariances between CIL and CT were freely estimated within each group to account for their conceptual and empirical relatedness.

Given the ordinal nature of the screen time variable, the model was estimated using the weighted least squares mean- and variance-adjusted (WLSMV) estimator, as implemented in the *lavaan* package in R. In *lavaan*, this estimator is operationalized as diagonally weighted least squares (DWLS) with robust standard errors and Wald-type test statistics. Under this framework, the ordinal screen time variable is modeled as a categorical indicator of an underlying latent response variable, allowing regression coefficients to be interpreted as effects on this latent propensity. The theta parameterization was used to enhance estimation stability. Under WLSMV, the ordinal $ST$ is modeled via a latent response variable $ST^*$:

$$ST^*_{ic} = d_c\, SES_{ic} + e_{ic}, \qquad e_{ic} \sim N(0,1), \tag{1}$$

$$ST_{ic} = \begin{cases} 1, & ST^*_{ic} \le \tau_{c,1}, \\ 2, & \tau_{c,1} < ST^*_{ic} \le \tau_{c,2}, \\ 3, & ST^*_{ic} > \tau_{c,2}. \end{cases} \tag{2}$$

Consequently, regression coefficients involving $ST$ should be interpreted as effects on the latent index $ST^{*}$.

The residual covariance between CIL and CT was positive in all three countries, consistent with the close relation between the two achievement outcomes after conditioning on SES, ST, and G. Let c ∈ {KR,TW,PT} index countries and let Y ∈ {CIL,CT} denote outcomes. The structural model is:

$$ST^{*}_{ic} = d_{c} SES_{ic} + \varepsilon^{(ST)}_{ic},$$
$$G_{ic} = a^{(G)}_{SES,c} SES_{ic} + a^{(G)}_{ST,c} ST^{*}_{ic} + \varepsilon^{(G)}_{ic},$$
$$Y_{ic} = c^{(Y)}_{SES,c} SES_{ic} + c^{(Y)}_{ST,c} ST^{*}_{ic} + b^{(Y)}_{G,c} G_{ic} + \varepsilon^{(Y)}_{ic}.$$

Cross-national differences were assessed through a series of model comparisons and parameter tests. First, a global test was conducted to examine whether structural paths differed significantly across countries. Subsequently, direct and indirect effects were evaluated within each country. Pairwise comparisons were then performed to identify specific differences in path coefficients between countries. This approach allows for a detailed examination of both overall and parameter-specific cross-group variation.

### 3.4. Plausible Value Combination

The ICILS dataset provides multiple plausible values (PVs) for each latent outcome, reflecting measurement uncertainty in the estimation of students' underlying proficiency constructs. Rather than representing precise individual scores, PVs are multiple imputations drawn from a posterior distribution of student ability, and are intended to yield unbiased population-level estimates (Mislevy et al., 1992; von Davier et al., 2009). Thus, we used all five PVs (PV1-PV5) for each outcome (CIL & CT). All model parameters were estimated separately for each PV and then combined using Rubin's rules (Rubin, 1987; Little & Rubin, 2019) to account for both within- and between-imputation variability. Let $m = 5$ denote the number of PVs. For each $l = 1, ..., 5$, let $\hat{\theta}^{(l)}$ denote the parameter estimate and $\hat{Var}^{(l)}$ its variance. The pooled estimate is

$$\bar{\theta} = \frac{1}{m} \sum_{l=1}^{m} \hat{\theta}^{(l)}.$$

The within-imputation variance is

$$W = \frac{1}{m} \sum_{l=1}^{m} \hat{Var}^{(l)},$$

and the between-imputation variance is

$$B = \frac{1}{m-1} \sum_{l=1}^{m} \left( \hat{\theta}^{(l)} - \bar{\theta} \right)^{2}.$$

The total variance is

$$T = W + \left(1 + \frac{1}{m}\right)B.$$

All reported standard errors are computed as $\sqrt{T}$, and statistical inference is based on these combined estimates.

This procedure ensures that statistical inferences properly reflect the uncertainty inherent in large-scale assessment data.

### 3.4. Conceptual Model

To examine the structural relationships among SES, screen time regulation, ICT self-efficacy, and digital outcomes, we specified a multigroup structural equation model. Figure 1 presents the conceptual model guiding the analysis. In the model, SES was treated as an exogenous variable predicting screen time, self-efficacy, and both outcome variables. Screen time was specified as an intermediate variable, predicted by SES and in turn predicting self-efficacy as well as the two outcomes. In addition, direct paths were specified from SES and screen time to both CIL and CT to account for potential direct effects beyond the mediated pathways. Both direct and indirect effects were examined to capture the complexity of relationships among SES, screen time, self-efficacy, and digital competence.

**Figure 1**
*Conceptual model*

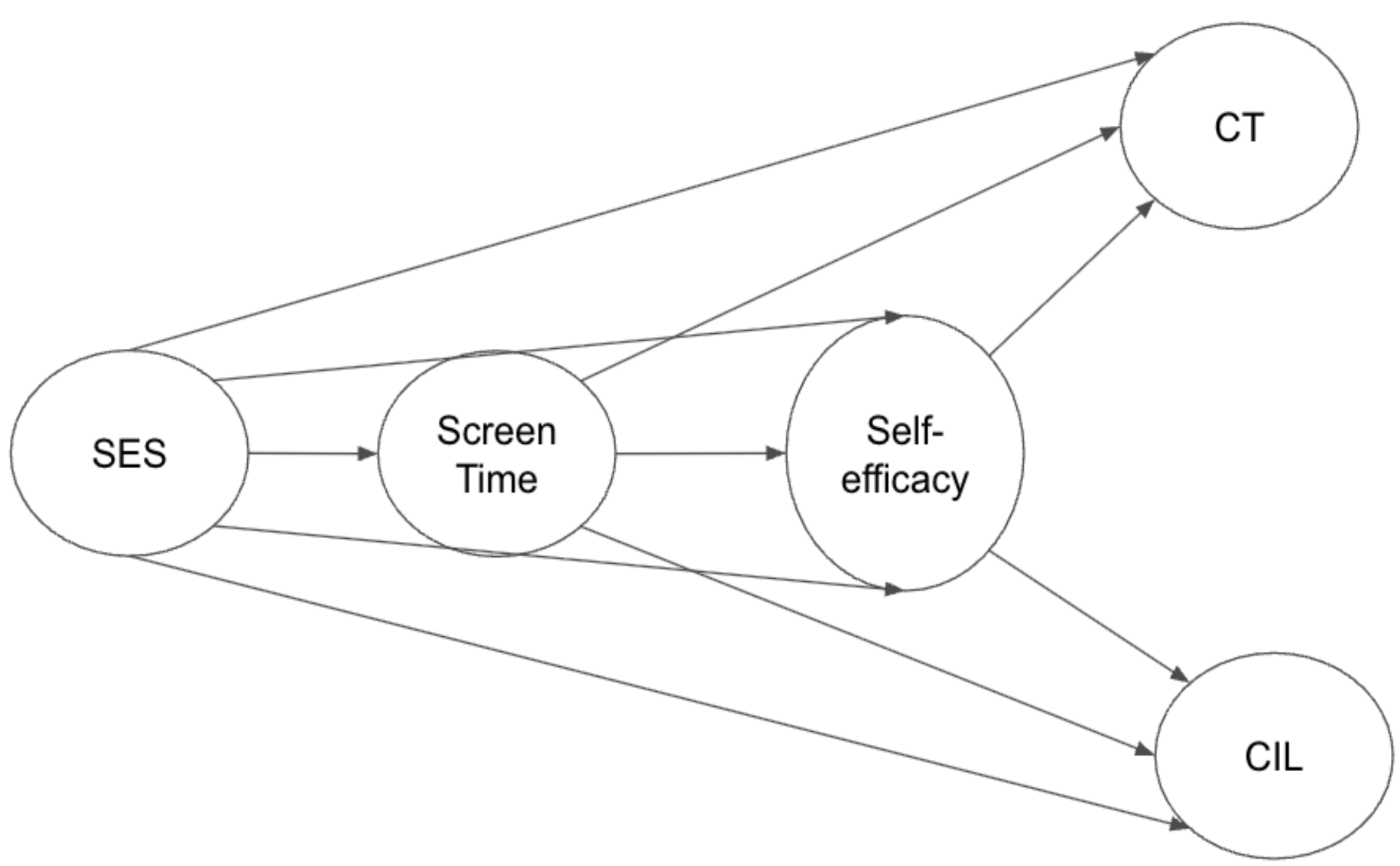

## 4. Results

**4.1. Cross-Country Heterogeneity in Structural Regression**

The multi-group ordinal SEM was estimated using WLSMV, which accommodates the ordinal screen time variable via an underlying latent response formulation. To formally assess whether the structural regression coefficients differ across countries, we conducted a global Wald-type test of the joint null hypothesis that all regression slopes are equal across groups, with inference combined across plausible values using Rubin's multiple-imputation framework.

For each plausible value, a robust Wald statistic is computed based on the sandwich covariance estimator under WLSMV. These test statistics are then combined across imputations using the D2 pooling method, which yields an approximate F-statistic accounting for both within-imputation and between-imputation variability.

In our analysis, the global pooled test was statistically significant, $F(18, 181.312) = 11.108$, $p < .001$. The average relative increase in variance (ARIV) was $.067$, and the fraction of missing information (FMI) was $.062$, indicating that plausible value uncertainty contributed modestly to the pooled Wald inference. This result indicates that at least one structural regression coefficient differs across countries. We therefore conclude that there is statistically detectable cross-country heterogeneity in the regression structure of the model.

To identify which specific structural regression coefficients differ across countries, we conducted per-path Wald-type tests under the WLSMV estimator, with inference combined across plausible values using the D2 pooling method. Table 1 summarizes per-path pooled Wald tests of cross-country equality in the model.

**Table 1**
*Path homogeneity test across countries*

| Path | $df$ | $p$ |
|---|---|---|
| SES → ST | 2 | $<.001$ |
| SES → G | 2 | $<.001$ |
| ST → G | 2 | $.004$ |
| SES → CIL | 2 | $<.001$ |
| ST → CIL | 2 | $<.001$ |
| G → CIL | 2 | $<.027$ |
| SES → CT | 2 | $.080$ |
| ST → CT | 2 | $<.001$ |
| G → CT | 2 | $<.001$ |

With the exception of $SES \rightarrow CT$ ($p =.080$), all paths show statistically significant cross-country heterogeneity. This means that the structural relations linking SES, screen time, self-efficacy, and the two outcomes are not invariant across countries. The strongest evidence of heterogeneity appears for the upstream SES-screen time relationship, the SES-G pathway, and the screen time-outcome pathways. The non-significant test for SES→ CT indicates that, despite numerical variation, the direct SES effect on CT is not statistically distinguishable across countries in this specification.

### 4.2. Structural Direct Path Estimates

Table 2 summarizes the primary structural coefficients for each group. Standard errors are robust and inference is based on Wald-type $z$ tests.

**Table 2**

*Structural direct path standardized coefficients*

| | KR | | | TW | | | PT | | |
|---|---|---|---|---|---|---|---|---|---|
| Path | Est. | SE | $p$ | Est. | SE | $p$ | Est. | SE | $p$ |
| $SES \rightarrow ST$ | $-0.176$ | $0.026$ | $<.001$ | $-0.303$ | $0.018$ | $<.001$ | $-0.053$ | $0.026$ | $.046$ |
| $SES \rightarrow G$ | $0.209$ | $0.020$ | $<.001$ | $0.160$ | $0.017$ | $<.001$ | $0.192$ | $0.020$ | $<.001$ |
| $ST \rightarrow G$ | $0.011$ | $0.027$ | $.675$ | $0.013$ | $0.020$ | $.524$ | $0.102$ | $0.026$ | $<.001$ |
| $SES \rightarrow CIL$ | $0.153$ | $0.019$ | $<.001$ | $0.301$ | $0.015$ | $<.001$ | $0.309$ | $0.018$ | $<.001$ |
| $ST \rightarrow CIL$ | $-0.024$ | $0.026$ | $.347$ | $0.003$ | $0.019$ | $.884$ | $0.126$ | $0.024$ | $<.001$ |
| $G \rightarrow CIL$ | $0.241$ | $0.017$ | $<.001$ | $0.196$ | $0.013$ | $<.001$ | $0.188$ | $0.017$ | $<.001$ |
| $SES \rightarrow CT$ | $0.158$ | $0.019$ | $<.001$ | $0.270$ | $0.016$ | $<.001$ | $0.305$ | $0.018$ | $<.001$ |
| $ST \rightarrow CT$ | $-0.071$ | $0.025$ | $.004$ | $-0.031$ | $0.019$ | $.107$ | $0.080$ | $0.025$ | $.001$ |
| $G \rightarrow CT$ | $0.163$ | $0.017$ | $<.001$ | $0.134$ | $0.014$ | $<.001$ | $0.091$ | $0.018$ | $<.001$ |

In Korea, SES is negatively associated with latent screen time ($\hat{\beta}_{std} =- 0.176$, $p <.001$), indicating that higher-SES students tend to exhibit lower latent screen time propensity. SES is also positively associated with general self-efficacy ($\hat{\beta}_{std} = 0.209, p <.001$). In contrast, $ST$ does not significantly predict $G$, indicating no meaningful screen time-to-self-efficacy pathway in Korea once SES is controlled. SES has positive direct effects on both CIL ($\hat{\beta}_{std} = 0.153, p <.001$) and CT ($\hat{\beta}_{std} = 0.158, p <.001$). The direct effect of ST on CIL is not significant, whereas its direct effect on CT is negative and significant ($\hat{\beta}_{std} =- 0.071, p =.004$). General self-efficacy is positively associated with both outcomes (CIL: $\hat{\beta}_{std} = 0.241$; CT: $\hat{\beta}_{std} = 0.163$, both $p <.001$). Thus, in Korea, achievement is positively tied to SES and self-efficacy, while screen time has a negative direct association only with CT.

In Taiwan, SES is strongly negatively associated with latent screen time ($\hat{\beta}_{std} =- 0.303$, $p <.001$), the strongest SES-to-ST association among the three countries. SES is positively associated with $G$ ($\hat{\beta}_{std} = 0.160$, $p < .001$), but ST does not significantly predict $G$. SES has strong positive direct effects on both CIL ($\hat{\beta}_{std} = 0.301$, $p < .001$) and CT ($\hat{\beta}_{std} = 0.270$, $p < .001$). In contrast, ST has no statistically significant direct effects on either CIL or CT. General self-efficacy is positively associated with both outcomes (CIL: $\hat{\beta}_{std} = 0.196$; CT: $\hat{\beta}_{std} = 0.134$, both $p < .001$). Thus, in Taiwan, SES and general self-efficacy are significantly associated with both outcomes, whereas screen time regulation shows no statistically significant direct or indirect associations.

In Portugal, SES is also negatively associated with latent screen time ($\hat{\beta}_{std} = -0.053$, $p = .046$), although the association is much smaller than in Korea or Taiwan. SES is positively associated with $G$ ($\hat{\beta}_{std} = 0.192$, $p < .001$). Unlike Korea and Taiwan, ST significantly and positively predicts $G$ in Portugal ($\hat{\beta}_{std} = 0.102$, $p < .001$). SES has positive direct effects on both CIL ($\hat{\beta}_{std} = 0.309$, $p < .001$) and CT ($\hat{\beta}_{std} = 0.305$, $p < .001$). Importantly, ST has positive direct effects on both CIL ($\hat{\beta}_{std} = 0.126$, $p < .001$) and CT ($\hat{\beta}_{std} = 0.080$, $p = .001$), in sharp contrast to the Korean and Taiwanese patterns. General self-efficacy is positively associated with both CIL ($\hat{\beta}_{std} = 0.188$, $p < .001$) and CT ($\hat{\beta}_{std} = 0.091$, $p < .001$). Thus, in Portugal, less restrictive screen time regulation is positively associated with both outcomes directly and indirectly through general self-efficacy.

### 4.3. Indirect Effects

Table 3 shows the estimated indirect effects from the multi-group ordinal SEM. For Korea, the indirect effect of SES on CIL through screen time is not statistically significant. The indirect effect of ST on CIL through $G$ is also not significant, consistent with the non-significant $ST \rightarrow G$ direct path. In contrast, the indirect effect of SES through $G$ is statistically significant for CIL ($\hat{\beta} = 5.349$, $p < .001$) and CT ($\hat{\beta} = 5.044$, $p < .001$). The serial pathway $SES \rightarrow ST \rightarrow G \rightarrow Y$ is not significant for either outcome. The total ST effect is negative and significant for CT ($\hat{\beta} = -7.987$, $p = .008$), but not for CIL.

In Taiwan, none of the indirect effects involving ST are statistically significant for either outcome. The indirect effect of SES through $G$ is significant for both CIL ($\hat{\beta} = 2.714$, $p < .001$) and CT ($\hat{\beta} = 2.316$, $p < .001$). These results indicate that achievement differences in Taiwan are carried primarily by SES and general self-efficacy, with negligible contribution from screen time.

Portugal shows a distinct pattern. The indirect effect of ST through $G$ is positive and significant for both CIL ($\hat{\beta} = 1.423$, $p < .001$) and CT ($\hat{\beta} = 0.833$, $p = .005$). The indirect effect of SES through $G$ is also significant for both outcomes (CIL: $\hat{\beta} = 2.407$; CT: $\hat{\beta} = 1.409$, both $p < .001$). However, the serial pathway $SES \rightarrow ST \rightarrow G \rightarrow Y$ remains non-significant. Overall, Portugal is the country in which the indirect effect of screen time regulation through general self-efficacy is statistically significant for both outcomes.

**Table 3**

*Indirect effects from the multi-group ordinal SEM*

| Outcome | Effect (country-specific) | Estimate | SE | $p$ |
|---|---|---|---|---|
| $CIL$ | $IE_{SES \rightarrow ST}$ (KR) | $0.452$ | $0.483$ | $.351$ |
| $CIL$ | $IE_{SES \rightarrow ST}$ (TW) | $-0.072$ | $0.496$ | $.885$ |
| $CIL$ | $IE_{SES \rightarrow ST}$ (PT) | $-0.444$ | $0.240$ | $.064$ |
| $CIL$ | $IE_{ST \rightarrow G}$ (KR) | $0.227$ | $0.524$ | $.664$ |
| $CIL$ | $IE_{ST \rightarrow G}$ (TW) | $0.206$ | $0.337$ | $.540$ |
| $CIL$ | $IE_{ST \rightarrow G}$ (PT) | $1.423$ | $0.368$ | $<.001$ |
| $CIL$ | $IE_{SES \rightarrow G}$ (KR) | $5.349$ | $0.637$ | $<.001$ |
| $CIL$ | $IE_{SES \rightarrow G}$ (TW) | $2.714$ | $0.355$ | $<.001$ |
| $CIL$ | $IE_{SES \rightarrow G}$ (PT) | $2.407$ | $0.331$ | $<.001$ |
| $CIL$ | $IE_{SES \rightarrow ST \rightarrow G}$ (KR) | $-0.051$ | $0.119$ | $.665$ |
| $CIL$ | $IE_{SES \rightarrow ST \rightarrow G}$ (TW) | $-0.066$ | $0.108$ | $.541$ |
| $CIL$ | $IE_{SES \rightarrow ST \rightarrow G}$ (PT) | $-0.068$ | $0.037$ | $.069$ |
| $CT$ | $IE_{SES \rightarrow ST}$ (KR) | $1.854$ | $0.690$ | $.009$ |
| $CT$ | $IE_{SES \rightarrow ST}$ (TW) | $1.022$ | $0.637$ | $.109$ |
| $CT$ | $IE_{SES \rightarrow ST}$ (PT) | $-0.342$ | $0.217$ | $.117$ |
| $CT$ | $IE_{ST \rightarrow G}$ (KR) | $0.214$ | $0.536$ | $.690$ |
| $CT$ | $IE_{ST \rightarrow G}$ (TW) | $0.176$ | $0.289$ | $.543$ |
| $CT$ | $IE_{ST \rightarrow G}$ (PT) | $0.833$ | $0.291$ | $.005$ |
| $CT$ | $IE_{SES \rightarrow G}$ (KR) | $5.044$ | $0.755$ | $<.001$ |
| $CT$ | $IE_{SES \rightarrow G}$ (TW) | $2.316$ | $0.355$ | $<.001$ |
| $CT$ | $IE_{SES \rightarrow G}$ (PT) | $1.409$ | $0.331$ | $<.001$ |

| Outcome | Effect (country-specific) | Estimate | SE | $p$ |
|---|---|---|---|---|
| $CT$ | $IE_{SES \to ST \to G}$ (KR) | $-0.048$ | $0.122$ | $.690$ |
| $CT$ | $IE_{SES \to ST \to G}$ (TW) | $-0.056$ | $0.093$ | $.543$ |
| $CT$ | $IE_{SES \to ST \to G}$ (PT) | $-0.040$ | $0.027$ | $.138$ |

### 4.4. Pairwise Country Contrasts for Structural Path Coefficients

We conducted pairwise country contrasts for structural path coefficients. Table 4 reports the results. These contrasts are computed as Group 1 minus Group 2 and should be interpreted as cross-country comparisons of structural paths, not as indirect or mediated effects.

**Table 4**

*Pairwise country contrasts for structural path coefficients*

| Path | Comparison | $\hat{\Delta}$ | SE | Lower | Upper | $p$ |
|---|---|---|---|---|---|---|
| G → CIL | KR vs PT | 0.543 | 0.198 | 0.152 | 0.934 | $.007$ |
| G → CIL | KR vs TW | 0.174 | 0.190 | -0.203 | 0.551 | $.362$ |
| G → CIL | TW vs PT | 0.369 | 0.184 | -0.004 | 0.743 | $.052$ |
| G → CT | KR vs PT | 1.037 | 0.267 | 0.511 | 1.560 | $<.001$ |
| G → CT | KR vs TW | 0.328 | 0.260 | -0.184 | 0.840 | $.208$ |
| G → CT | TW vs PT | 0.709 | 0.238 | 0.231 | 1.190 | $.005$ |
| SES → CIL | KR vs PT | -4.457 | 2.440 | -9.290 | 0.378 | $.070$ |
| SES → CIL | KR vs TW | -9.853 | 2.490 | -14.700 | -4.960 | $<.001$ |
| SES → CIL | TW vs PT | 5.397 | 1.950 | 1.550 | 9.240 | $.006$ |
| SES → CT | KR vs PT | -1.464 | 3.310 | -7.970 | 5.040 | $.659$ |
| SES → CT | KR vs TW | -5.992 | 3.430 | -12.700 | 0.736 | $.081$ |
| SES → CT | TW vs PT | 4.528 | 2.430 | -0.282 | 9.340 | $.065$ |
| SES → G | KR vs PT | 1.020 | 0.315 | 0.403 | 1.640 | $.001$ |
| SES → G | KR vs TW | 1.183 | 0.308 | 0.580 | 1.790 | $<.001$ |
| SES → G | TW vs PT | -0.163 | 0.236 | -0.626 | 0.300 | $.490$ |
| SES → ST | KR vs PT | -0.179 | 0.042 | -0.261 | -0.096 | $<.001$ |
| SES → ST | KR vs TW | 0.094 | 0.040 | 0.015 | 0.173 | $.019$ |
| SES → ST | TW vs PT | -0.273 | 0.031 | -0.334 | -0.211 | $<.001$ |
| ST → CIL | KR vs PT | -11.351 | 2.780 | -16.800 | -5.910 | $<.001$ |
| ST → CIL | KR vs TW | -2.223 | 2.630 | -7.440 | 2.990 | $.400$ |
| ST → CIL | TW vs PT | -9.128 | 2.360 | -13.800 | -4.440 | $<.001$ |
| ST → CT | KR vs PT | -15.396 | 3.620 | -22.500 | -8.250 | $<.001$ |
| ST → CT | KR vs TW | -5.007 | 3.480 | -12.000 | 1.940 | $.155$ |

| Path | Comparison | $\hat{\Delta}$ | SE | Lower | Upper | $p$ |
|---|---|---|---|---|---|---|
| ST → CT | TW vs PT | -10.390 | 2.980 | -16.300 | -4.490 | $<.001$ |
| ST → G | KR vs PT | -0.857 | 0.368 | -1.580 | -0.136 | $.020$ |
| ST → G | KR vs TW | 0.001 | 0.322 | -0.630 | 0.631 | $.998$ |
| ST → G | TW vs PT | -0.857 | 0.307 | -1.460 | -0.256 | $.005$ |

Several systematic cross-country differences emerge from Table 4. First, the association between SES and screen time regulation differs significantly across all three countries. It is strongest and most negative in Taiwan, followed by Korea, and weakest in Portugal. Second, the associations between screen time regulation and both digital outcomes are significantly larger in Portugal than in Korea and Taiwan, whereas Korea and Taiwan do not differ significantly. A similar pattern is observed for the association between screen time regulation and general self-efficacy. Third, the unstandardized association between SES and CIL is significantly larger in Taiwan than in Korea and Portugal, whereas Korea and Portugal do not differ significantly. For CT, none of the pairwise SES comparisons reach statistical significance. Finally, the association between self-efficacy and digital outcomes are weakest in Portugal, especially for CT. Self-efficacy remains beneficial in all countries, but its outcome association is smaller in Portugal than in Korea and Taiwan.

**4.5. Pairwise Country Contrasts for Indirect Effects**

Indirect effects are functions of the estimated structural coefficients and quantify specific mediated pathways. For each indirect effect and each pair of countries, we computed the contrast $\Delta_{g_1,g_2} := IE_{g_1} - IE_{g_2}$, with standard errors and confidence intervals obtained from the pooled covariance matrix using the delta method. A positive contrast indicates that the corresponding indirect effect is larger in Group 1 than in Group 2, whereas a negative contrast indicates that it is smaller. The resulting pairwise contrasts are reported in Table 5.

**Table 5**

*Pairwise country contrasts for indirect effects*

| Outcome | Effect | Comparison | $\hat{\Delta}$ | SE | Lower | Upper | $p$ |
|---|---|---|---|---|---|---|---|
| CIL | $\mathrm{IE}_{SES\to G}$ | KR vs PT | 2.942 | 0.718 | 1.534 | 4.350 | $<.001$ |
| CIL | $\mathrm{IE}_{SES\to G}$ | KR vs TW | 2.635 | 0.729 | 1.205 | 4.066 | $<.001$ |
| CIL | $\mathrm{IE}_{SES\to G}$ | TW vs PT | 0.307 | 0.485 | -0.651 | 1.264 | $.528$ |
| CIL | $\mathrm{IE}_{SES\to ST}$ | KR vs PT | 0.896 | 0.539 | -0.164 | 1.955 | $.097$ |
| CIL | $\mathrm{IE}_{SES\to ST}$ | KR vs TW | 0.524 | 0.692 | -0.849 | 1.896 | $.451$ |

| Outcome | Effect | Comparison | $\hat{\Delta}$ | SE | Lower | Upper | $p$ |
|---|---|---|---|---|---|---|---|
| CIL | $IE_{SES \to ST}$ | TW vs PT | 0.372 | 0.551 | -0.714 | 1.458 | $.500$ |
| CIL | $IE_{SES \to ST \to G}$ | KR vs PT | 0.016 | 0.124 | -0.228 | 0.260 | $.896$ |
| CIL | $IE_{SES \to ST \to G}$ | KR vs TW | 0.015 | 0.160 | -0.300 | 0.329 | $.927$ |
| CIL | $IE_{SES \to ST \to G}$ | TW vs PT | 0.002 | 0.114 | -0.222 | 0.225 | $.989$ |
| CIL | $IE_{ST \to G}$ | KR vs PT | -1.196 | 0.640 | -2.450 | 0.059 | $.062$ |
| CIL | $IE_{ST \to G}$ | KR vs TW | 0.021 | 0.623 | -1.200 | 1.242 | $.973$ |
| CIL | $IE_{ST \to G}$ | TW vs PT | -1.217 | 0.499 | -2.195 | -0.239 | $.015$ |
| CT | $IE_{SES \to G}$ | KR vs PT | 3.634 | 0.824 | 2.015 | 5.253 | $<.001$ |
| CT | $IE_{SES \to G}$ | KR vs TW | 2.727 | 0.834 | 1.091 | 4.363 | $.001$ |
| CT | $IE_{SES \to G}$ | TW vs PT | 0.907 | 0.485 | -0.055 | 1.869 | $.064$ |
| CT | $IE_{SES \to ST}$ | KR vs PT | 2.196 | 0.724 | 0.759 | 3.632 | $.003$ |
| CT | $IE_{SES \to ST}$ | KR vs TW | 0.832 | 0.939 | -1.033 | 2.697 | $.378$ |
| CT | $IE_{SES \to ST}$ | TW vs PT | 1.364 | 0.673 | 0.043 | 2.685 | $.043$ |
| CT | $IE_{SES \to ST \to G}$ | KR vs PT | -0.009 | 0.124 | -0.253 | 0.235 | $.943$ |
| CT | $IE_{SES \to ST \to G}$ | KR vs TW | 0.008 | 0.153 | -0.292 | 0.307 | $.959$ |
| CT | $IE_{SES \to ST \to G}$ | TW vs PT | -0.017 | 0.096 | -0.206 | 0.172 | $.862$ |
| CT | $IE_{ST \to G}$ | KR vs PT | -0.619 | 0.610 | -1.815 | 0.578 | $.311$ |
| CT | $IE_{ST \to G}$ | KR vs TW | 0.038 | 0.609 | -1.156 | 1.233 | $.950$ |
| CT | $IE_{ST \to G}$ | TW vs PT | -0.657 | 0.410 | -1.462 | 0.148 | $.109$ |

Table 5 shows that heterogeneity in indirect effects is selective rather than universal. For $CIL$, the indirect pathway $SES \to G \to CIL$ is significantly larger in KR than in both PT ( $\hat{\Delta} = 2.942, p < .001$) and TW ($\hat{\Delta} = 2.635, p < .001$), whereas TW and PT do not differ. Thus, the SES-to-G-to-CIL pathway is strongest in Korea. The pathway $SES \to ST \to CIL$ does not show significant pairwise differences, and the serial pathway $SES \to ST \to G \to CIL$ is stable across countries. For $ST \to G \to CIL$, the TW-PT contrast is significant ($\hat{\Delta} = -1.217, p = .015$ ), indicating that this indirect pathway is larger in PT than in TW; the KR-PT contrast is marginal.

For $CT$, the indirect pathway $SES \rightarrow G \rightarrow CT$ is again larger in Korea than in both Portugal ($\hat{\Delta} = 3.634, p < .001$) and Taiwan ($\hat{\Delta} = 2.727, p = .001$), while Taiwan and Portugal are only marginally different. The pathway $SES \rightarrow ST \rightarrow CT$ shows significant differences for KR-PT ($\hat{\Delta} = 2.196, p = .003$) and TW-PT ($\hat{\Delta} = 1.364, p = .043$), indicating that this screen time-mediated SES pathway is weaker in Portugal than in Korea and Taiwan. As with CIL, the serial pathway $SES \rightarrow ST \rightarrow G \rightarrow CT$ shows no significant cross-country differences. Finally, none of the pairwise contrasts for $ST \rightarrow G \rightarrow CT$ reach conventional significance, although the pattern is again numerically stronger in Portugal than in Taiwan.

Overall, the indirect effect comparisons indicate two main sources of cross-country heterogeneity in the model. First, the SES-self-efficacy-digital outcome pathway is strongest in Korea. Second, the ST-self-efficacy-digital outcome mechanism is most evident in Portugal, especially for CIL. In contrast, the higher-order serial indirect pathway $SES \rightarrow ST \rightarrow G \rightarrow Y$ is small and statistically stable across countries for both outcomes.

## 5. Discussion

The present study examined the structural relationships between SES, screen time regulation, ICT self-efficacy, and empirical digital outcomes across countries. Our findings reveal substantial cross-national heterogeneity in the structural pathways, directly responding to international calls to explore the highly contextualized nature of youth digital engagement.

### 5.1. Screen Time and Technological Affordances

The role of screen time regulation was inconsistent and varied substantially across countries. A key empirical contribution of this study is unpacking the distinct direct trajectories of screen time across the three contexts. In Taiwan, screen time demonstrated no significant direct or indirect associations with either CIL or CT. In South Korea, screen time exerted a selective, negative direct effect specifically on CT. Conversely, in Portugal, screen time emerged as a robust, positive predictor of both CIL and CT.

Because our empirical model utilizes student-reported data on screen time limits, these parameters fundamentally capture household-level variation rather than macro-level mandates. It is critical to note that screen time regulation is highly idiosyncratic and varies within any single country. Nevertheless, the aggregate parameter differences suggest that the functional nature of screen engagement differs across these systems (Fraillon, 2025; Kim et al., 2025; Pizarro et al., 2023). In South Korea and Taiwan, the strong negative association between SES and screen time indicates that socioeconomically advantaged families actively curtail device duration, potentially prioritizing traditional, high-stakes academic preparation (Kim et al., 2025). In such environments, when screen usage does occur, it may primarily represent recreational or passive consumption. This supports the displacement hypothesis where passive screen exposure potentially displaces deep, cognitively demanding tasks required for computational thinking.

Conversely, the Portuguese results demonstrate a unique pattern where cumulative screen time serves as a positive affordance. Rather than representing a direct consequence of Portugal's flexible national public health guidelines (Pizarro et al., 2023), this quantitative trend suggests that families in this context may lean toward active mediation rather than restrictive prohibition. Under a more permissive or collaborative domestic framework, longer screen durations likely capture active exploration, autonomous problem-solving, and digital creation, confirming that the quality of digital engagement shapes cognitive returns far more than simple duration metrics.

These findings reinforce prior literature suggesting that the relationship between screen-based activities and learning outcomes is highly inconsistent (Orben & Przybylski, 2019; Adelantado-Renau et al., 2019) and cannot be adequately explained by examining screen time in isolation. More importantly, this study advances the literature by demonstrating that these inconsistencies are not random but structurally embedded. Rather than functioning as an independent predictor, screen time operates within a broader system of relationships shaped by SES and self-efficacy. In particular, the results highlight that screen time becomes meaningful only when considered in relation to students' psychological engagement with technology. This emphasizes that the educational impact of technology depends more on how it is used than on how much it is used.

### 5.2. Self-Efficacy and Digital Competences

ICT self-efficacy emerged as a consistent and robust predictor of digital outcomes across all countries. Regardless of national context, students with higher levels of ICT self-efficacy demonstrated significantly stronger performance in both CIL and CT. These findings are consistent with a substantial body of literature highlighting the central role of self-efficacy in shaping learning behaviors and outcomes (Bandura, 1997; Hatlevik et al., 2018; Scherer & Siddiq, 2019). Importantly, the results further suggest that self-efficacy functions as a key mechanism through which structural and behavioral factors influence digital competence. For example, in Portugal, screen time was positively associated with self-efficacy, which in turn contributed to higher digital outcomes. In contrast, this pathway was not observed in Korea or Taiwan, indicating that similar behavioral patterns may translate into different psychological experiences across contexts.

These findings provide a critical theoretical contribution by shifting the focus from quantity-based measures of digital engagement to quality-oriented and psychologically mediated perspectives. While screen time has often been used as a convenient proxy for digital exposure, the present results suggest that such measures are insufficient for capturing meaningful engagement. Instead, the quality of interaction with technology, reflected in students' confidence, agency, and purposeful use, appears to be a more salient determinant of digital learning outcomes (Hernandez et al., 2026; Lee et al., 2026; Lee & Park, 2024). Thus, educators need to emphasize the importance of meaningful and functional technology use over frequency of use.

### 5.3. Socioeconomic Anchors and the Digital Divide

Beyond immediate behavioral and psychological variations, the multi-group analysis firmly establishes SES as an omnipresent structural anchor of the contemporary digital divide. Across all three national cohorts, the structural models reveal that higher SES consistently and significantly predicts superior digital outcomes for both CIL and CT. This universal resilience solidifies the consensus within literature that socioeconomic capital remains a primary determinant of student digital competence, granting high-SES students' greater access to enriched learning resources and structured academic support (Fraillon, 2025; Scherer & Siddiq, 2019).

At the psychological level, SES emerged as a universal, positive predictor of ICT self-efficacy across all three cohorts. These pathways corroborate the theoretical premise that parental education and occupational status do not merely secure physical access to premium hardware; rather, they fundamentally shape the psychological posture of the learner. These findings suggest that high socioeconomic environments contribute to the development of efficacy beliefs by shaping students' access to mastery experiences and supportive learning contexts. Importantly, our results indicate that the reproduction of socioeconomic advantage does not merely operate as an external direct force; it simultaneously conditions the structural and psychological landscapes of the household. At the behavioral level, SES demonstrated a negative association with screen time across all three countries, and this trajectory was heavily pronounced in Taiwan and South Korea relative to Portugal. This distinct implies that high-SES families actively enforce a more rigid, restrictive digital mediation framework. What the present study adds is a more nuanced account of how SES is embedded within a broader structural system. SES was not merely a background correlate of achievement; it was also linked to students' digital beliefs and patterns of digital engagement, and those links differed across national contexts. This underscores the importance of conceptualizing digital inequality not simply as unequal access or unequal performance, but as a layered process in which socioeconomic resources shape the motivational and experiential conditions under which digital competencies develop.

### 5.4. Theoretical Contributions and Practical Implications

By synthesis of the multi-group results, this study offers distinct contributions to the theoretical and practical landscapes of educational technology. First, this study directly addresses a persistent literature gap by empirical resolution of *why* international benchmarks like ICILS capture conflicting trends in student screen time. By demonstrating that the direct path from screen time regulation to digital competence, we provide structural proof that screen time cannot be conceptualized as a monolithic variable. Second, we extend social cognitive theory within cross-cultural contexts by revealing that the development of self-efficacy is structurally dependent on domestic constraints. While self-efficacy universally drives digital outcomes, its cultivation through technology exposure is only activated in permissive domestic environments, offering a more nuanced framework for future mediation models.

From a practical standpoint, educational stakeholders and policymakers must transcend simplistic, reductionist narratives regarding uniform screen limits for youth. Blanket time constraints that treat all screen hours equally fail to protect digital skill acquisition, as they risk discouraging the autonomous exploration necessary for high-level computational development. Furthermore, given that SES primarily transmits its advantages through the psychological channel of self-efficacy rather than the mere physical restriction of devices, interventions must look beyond hardware distribution. Schools and community programs serving lower-SES populations should deliberately design mastery experiences that build technological confidence and proactive digital agency, rather than merely mandating passive compliance or device exposure.

### 5.5. Limitations

While this study offers valuable insights into cross-national variations in youth digital literacy, several limitations must be addressed to preserve empirical objectivity. First, the analyses were based on cross-sectional ICILS data, so the findings should not be interpreted causally. The model identifies structural associations, not developmental effects over time. Second, the screen time indicator used here is necessarily limited in scope and does not capture the content, quality, or purpose of students' digital activity. The student-reported measures of screen time restriction inherently capture subjective perceptions and may be susceptible to social desirability or recall bias. Third, macro-level national digital policies and broad cultural frameworks were not directly measured as structural variables within our SEM framework. While our multi-group analysis successfully identified statistically significant parameter heterogeneity across countries, we cannot establish a direct causal link between national policy structures and individual household behaviors.

## 6. Conclusion

The present study offers an important contribution by demonstrating that the relationships among screen time, ICT self-efficacy, and digital outcomes are neither simple nor invariant across countries. Moving beyond descriptive comparisons, the findings demonstrate that the effects of screen time are contingent on broader socioeconomic and psychological contexts. Across three countries, screen time regulation showed inconsistent and context-dependent associations with digital outcomes. In contrast, ICT self-efficacy emerged as a consistent and powerful predictor of digital competence, highlighting the central role of psychological factors in shaping students' interactions with technology.

By integrating socioeconomic, behavioral, and psychological variables within a unified structural model, this study advances a more comprehensive understanding of digital learning.

The findings suggest that improving digital competence requires not only access to technology but also meaningful engagement and the development of confidence in using digital tools. Ultimately, this study responds directly to the research gap identified in ICILS by demonstrating that cross-national differences in digital outcomes can be better understood through the interplay of structural and psychological mechanisms. In doing so, it offers both theoretical and practical insights for advancing equitable and effective digital education in diverse contexts.